\documentstyle[11pt]{article}

\newtheorem{theorem}{{Theorem}}[section]
\newcommand{\TH}{\begin{theorem}.\, \ }
\newcommand{\THF}{\end{theorem}}

\title
{{\hfill\small\tt Theor.\,Math.\,Phys.\,115,\,737-739\,(1998)}
\\{\hfill}\\
Diophantine equations related to quasicrystals: a note}

\date{}
\author{{\footnotesize E. Pelantov\'a}\\
{\footnotesize\em Department of Mathematics}\\
{\footnotesize\em
Faculty of Nuclear Sciences and Physical Engineering, Prague,
Czech Republic}\\
{\footnotesize and}\\
 {\footnotesize  A.M. Perelomov\, \footnote{\, On
leave of absence from Institute of Theoretical and Experimental
Physics, 117259 Moscow, Russia}
}\\
{\footnotesize\em Departamento de F\'{i}sica Te\'orica,
Universidad de Zaragoza, Zaragoza, Spain }}

\begin{document}

\maketitle

\begin{abstract}\noindent
We give the general solution of three Diophantine equations in the
ring of integer of the algebraic number field ${\bf Q}[\sqrt{5}]$.
These equations are related to the problem of determination 
of the minimum distance in quasicrystals with fivefold symmetry.
\end{abstract}

\bigskip

\noindent{\bf 1. Introduction}\\[1mm]

Let us consider the ring of integers of algebraic number field
${\bf Q}[\sqrt{5}]$. This ring has a form \cite{Has}
\[
{\bf Z}[\tau] \equiv \{ m+n\tau \ |\ m,n \in {\bf Z}\ \},\] where
$\tau = (1+\sqrt{5})/2$, i.e.  ${\tau}^2=\tau +1$.

${\bf Z}[\tau]$-lattices are the basic structures used in the
theory of quasicrystals with fivefold symmetry. It is shown in
\cite{nas}, that problem introduced in [3] of determination of the
minimum distance in quasicrystals  for dimension $d=2,3,4$\ can be
transformed into the problem of finding the minimum value of
certain step function over the set $S_d$. For $d=2,3,4$, the
functions $S_d$ are defined as
\[ \begin{array}{lll}
S_2&=&\{(k_1,k_2)\in ({\bf Z}[\tau])^2\ |\ k_1^2+k_2^2-\tau
k_1k_2=1\}, \\
S_3&=&\{(k_1,k_2,k_3)\in ({\bf Z}[\tau])^3\ |\ k_1^2+k_2^2
+k_3^2-k_1k_2 -\tau k_2k_3=1\}, \\
S_4&=&\{(k_1,k_2,k_3,k_4)\in ({\bf Z}[\tau])^4\ |\ k_1^2+k_2^2
+k_3^2+k_4^2-k_1k_2-k_2k_3-\tau k_3k_4=1\}.
\end{array} \]
We find the general solutions of the Diophantine equations
describing $S_d$.

\bigskip

\noindent{\bf 2. The equation} \qquad $k_1^2+k_2^2-\tau k_1k_2=1$

\medskip

Let us note that the set of solutions $(k_1,k_2)$\ is invariant
under the transformations
\[
\Phi _1 (k_1,k_2)= (k_2,k_1)\ \ \ {\rm and}\ \ \ \Phi _2(k_1,k_2)=
(\tau k_2-k_1,k_2). \]
It is easy to see that  $\Phi _1^2=\Phi _2^2=(\Phi _1\Phi _2)^{10}=1$\ and
the group $G$\ generated by $ \Phi _1$\ and $ \Phi _2$\ has 20 elements.

Starting from the obvious solution $(1,0)$\ and using the
transformations of $G$\ we obtain ten solutions:
\begin{equation}\label{vsechno}
\begin{array}{ccccccccc}
(1,0)&\stackrel{\Phi _1 }{\longrightarrow}&
(0,1)&\stackrel{\Phi _2 }{\longrightarrow}&
(\tau,1)&\stackrel{\Phi _1 }{\longrightarrow}&
(1,\tau)&\stackrel{\Phi _2 }{\longrightarrow}&
(\tau, \tau)\\
\downarrow {\scriptstyle \Phi _2}&&&&&&&&\\
(-1,0)&\stackrel{\Phi _1 }{\longrightarrow}& (0,-1)&\stackrel{\Phi
_2 }{\longrightarrow}& (-\tau,-1)&\stackrel{\Phi _1
}{\longrightarrow}& (-1,-\tau)&\stackrel{\Phi _2
}{\longrightarrow}& (-\tau, -\tau).
\end{array}
\end{equation}

Substituting $k_j=a_j+\tau b_j,\ a_j,b_j\in {\bf Z}$\ for $j=1,2$ into
$f(k_1,k_2)=k_1^2+k_2^2-\tau k_1k_2=1$\ we obtain $f=A+B \tau =1+0\tau$\
and thus, in fact, the system of two Diophantine equations in ${\bf Z}$\
\begin{eqnarray}
A&\equiv &a_1^2+a_2^2 +b_1^2+b_2^2-b_1b_2-a_1b_2-a_2b_1=1,\\
B&\equiv &b_1^2+b_2^2+2(a_1b_1+a_2b_2)-2b_1b_2-a_1a_2-a_1b_2-a_2b_1 =0.
\end{eqnarray}
Rewriting (2) in the form
\begin{equation}\label{dvakrat}
a_1^2+a_2^2+(b_1-b_2)^2+(a_1-b_2)^2+(a_2-b_1)^2=2\ ,
\end{equation}
we can see that $a_j$\ may take only the three values $\pm 1$\ and
0. It follows immediately from (4) and (3) that there is no
solution with $a_1a_2\neq 0 $\ is absent and all other solutions
are
just the solutions given above. We have thus proved the following theorem.\\[2mm]
{\bf Theorem.}\ \ {\em The equation
\[ k_1^2+k_2^2-\tau k_1k_2=1 \]
for two variables $k_1,k_2\in {\bf Z}[\tau ]$\ has only the ten
solutions described in} (\ref{vsechno}).
\bigskip

\noindent {\bf 3. The equation} \qquad $k_1^2+k_2^2 +k_3^2-
k_1k_2-\tau k_2k_3=1$

\medskip
Substituting $k_j=a_j+ \tau b_j,\ a_j,b_j\in {\bf Z}$\ for
$j=1,2,3,$\ we obtain the system of two Diophantine equations
\begin{equation}\label{trojka}
a_1^2+a_2^2+a_3^2+b_1^2+b_2^2+b_3^2-a_1a_2-b_1b_2-a_2b_3-a_3b_2-b_2b_3=1,
\end{equation}
\begin{eqnarray}
\nonumber b_1^2+b_2^2+b_3^2 +2(a_1b_1+a_2b_2+a_3b_3-b_2b_3)-
~~~~~~~~~~~~~~~~~~~& {}&\\
~~~~~~~~~~~~~~~~~~~-a_1b_2-a_2b_1-b_1b_2 -a_2a_3-a_2b_3-a_3b_2& =&0.
\end{eqnarray}
Multiplying Eq.\,(\ref{trojka}) by 12 and rearranging it, we
obtain sum of squares in the left-hand side
\begin{eqnarray}\label{finit}
\nonumber 3(2a_1-a_2)^2+3(2b_1-b_2)^2+3(2a_3-b_2)^2+~~~~~~~~~~~~~~~~&{}&\\
~~~~~~~~~~~~~~~~~~~+6(b_3-b_2)^2+2(2a_2-b_3)^2+
(a_2-2b_3)^2&=&12.
\end{eqnarray}
It is now clear that the set of solutions is finite. To find all
of them, we study the inequality in two variables
\begin{equation}\label{dva}
2(2a_2-b_3)^2+(a_2-2b_3)^2\leq 12\ {\rm or}\,\, a_2^2 + 2
(b_3-a_2)^2 \leq 4
\end{equation}
(the left-hand side of the inequality is just the sum of the last
two squares in (\ref{finit})).

The only pairs $(a_2,b_3)$\  satisfying (\ref{dva}) are
\[ \pm (1,1)\ ,\ \  \pm (1,2),\, \ \pm (1,0)\ ,\ \ \pm (0,1)\ ,\ \ \pm (2,2)\ \
{\rm and}\ \ (0,0). \] We can restrict ourselves to the pairs with
$a_2+b_3\geq 0$\ because the set of solutions is centrally
symmetric. After a simple, but lengthy, calculation, we obtain the
corresponding solutions $(k_1,k_2,k_3)\in S_3 .$

For the pair
$(a_2,b_3)= \pm (1,1)$,\, these are \\
\centerline{$\pm(\tau ,\tau ^2 ,\tau )\ ,\ \ \pm(\tau ,\tau ^2
,\tau ^2)\ ,\ \ \pm(0,1,\tau )\ ,\ \ \pm(1,1,\tau )\ , \ \ \pm(1,
\tau ^2 ,\tau ),\ \ {\rm and}\ \  \pm(1, \tau ^2 ,\tau ^2)\ .$}
For the pair
$(a_2,b_3)= \pm (1,0)$,\, the corresponding solutions are \\
\centerline{$\pm(0,1,0)\ \ {\rm and}\ \ \pm(1,1,0)\ .$} For the
pair
$(a_2,b_3)= \pm (0,1)$,\ the corresponding solutions are \\
\centerline{$ \pm (\tau ,2\tau ,\tau ^2 )\ ,\ \ \pm (\tau ,\tau
,\tau )\ \ {\rm and} \ \ \pm (0,\tau ,\tau)\ .$}
For the pairs $(a_2,b_3)= \pm (2,2)$\ {\rm and} $(a_2,b_3)=\pm(1,2)$
, no solutions exist.\\[1mm]
For the pair
$(a_2,b_3)= (0,0)$,\, the corresponding solutions are \\
\centerline{$\pm(1,0,0)\ ,\ \ \pm(0,0,1)\ ,\ \ \pm(0,\tau ,1)\ \ {\rm and}\ \
\pm(\tau ,\tau ,1)\ .$}
\bigskip
\noindent {\em Thus, the set $S_3$\ has exactly 30 elements.}

\bigskip

\noindent{\bf 4. The equation} \qquad $k_1^2+k_2^2 +k_3^2+k_4^2 -
k_1k_2- k_2k_3 -\tau k_3k_4=1$

\medskip
Substituting $k_j=a_j+ \tau b_j,\ a_j,b_j\in {\bf Z}$\ for
$j=1,2,3,4$,\ gives the equations
\begin{eqnarray}\label{ctyrka}
\nonumber a_1^2 +b_1^2+a_2^2+b_2^2+a_3^2+b_3^2 +a_4^2+b_4^2-
~~~~~~~~~~~~~~~~~~~~~~~~~~~~~~~~~~~& {}&\\
~~~~~~~~~~~~~~~~~~~-a_1a_2-b_1b_2-a_2a_3
-b_2b_3-b_3a_4-a_3b_4-b_3b_4 & =&1
\end{eqnarray}
and
\begin{equation}
{\sum}_{i=1}^{4}b_i^2 + 2{\sum}_{i=1}^{4}a_ib_i - {\sum}_{i=1}^{3}
(a_ib_{i+1}+a_{i+1}b_i+b_ib_{i+1}) - a_3a_4-b_3b_4 = 0.
\end{equation}
The equation (\ref{ctyrka}) is equivalent to
\begin{eqnarray}\label{uprava}
\nonumber 4(2a_1-a_2)^2+4(2a_3-a_2-b_4)^2 +4(2b_3-b_2-b_4-a_4)^2
+a_4^2+&{}&\\
+4(2b_1-b_2)^2++2(2a_2-b_4)^2+2(2b_2-b_4-a_4)^2
+(2b_4-3a_4)^2  &=&16.
 \end{eqnarray}
 \noindent
{\em After a trivial, but tedious, calculation, we obtain 120 solutions.}\\[3mm]

 \noindent{\bf 5.\,\,Conclusion}

All solutions that we found correspond to the vectors of
 non-crystallographic root system (see \cite{CMP}).
 It was clear a priori that $S_d$\ must have these elements.
 Our main result is that no other solutions exist.

This work (Pelantov\'a) was partially supported by the Czech
Republic (Grant No. GA$\check{C}$R 202/97/0218).

\end{document}